\documentclass[twocolumn]{aastex62}
\usepackage{amsmath}
\usepackage{wasysym}      
\usepackage{graphicx}
\usepackage{amssymb}
\usepackage{epstopdf}
\usepackage{mathrsfs}
\usepackage{anyfontsize}
\usepackage{natbib}
\usepackage{color}
\usepackage{lipsum}
\usepackage{diagbox}

\newcommand{\Msun}{M_{\odot}}

\shorttitle{The Peak of the Fallback Rate from TDEs}
\shortauthors{Bandopadhyay et al.}
\begin{document}
\title{The Peak of the Fallback Rate from Tidal Disruption Events: Dependence on Stellar Type}
\author[0000-0002-5116-844X]{Ananya Bandopadhyay}
\affiliation{Department of Physics, Syracuse University, Syracuse, NY 13210, USA}
\author[0009-0008-5847-9778]{Julia Fancher}
\affiliation{Department of Physics, Syracuse University, Syracuse, NY 13210, USA}
\author{Aluel Athian}
\affiliation{Colby College, Waterville, ME 04901, USA}
\affiliation{Henninger High School, Syracuse, NY 13206, USA}
\author{Valentino Indelicato}
\affiliation{Nottingham High School, Syracuse, NY 13224, USA}
\author{Sarah Kapalanga}
\affiliation{University of Rochester, Rochester, NY 14627, USA}
\affiliation{Institute of Technology at Syracuse Central High School, Syracuse, NY 13202, USA}
\author{Angela Kumah}
\affiliation{Dartmouth College, Hanover, NH 03755, USA}
\affiliation{Nottingham High School, Syracuse, NY 13224, USA}
\author[0009-0003-8285-0702]{Daniel A.~Paradiso}
\affiliation{Department of Physics, Syracuse University, Syracuse, NY 13210, USA}
\author[0009-0007-3017-2195]{Matthew Todd}
\affiliation{Department of Physics, Syracuse University, Syracuse, NY 13210, USA}
\author[0000-0003-3765-6401]{Eric R.~Coughlin}
\affiliation{Department of Physics, Syracuse University, Syracuse, NY 13210, USA}
\author[0000-0002-2137-4146]{C.~J.~Nixon}
\affiliation{School of Physics and Astronomy, Sir William Henry Bragg Building, Woodhouse Ln., University of Leeds, Leeds LS2 9JT, UK}

\email{abandopa@syr.edu}
\email{ecoughli@syr.edu}

\begin{abstract}
A star completely destroyed in a tidal disruption event (TDE) ignites a luminous flare that is powered by the fallback of tidally stripped debris to a supermassive black hole (SMBH) of mass $M_{\bullet}$. We analyze two estimates for the peak fallback rate in a TDE, one being the ``frozen-in'' model, which predicts a strong dependence of the time to peak fallback rate, $t_{\rm peak}$, on both stellar mass and age, with $15\textrm{ days} \lesssim t_{\rm peak} \lesssim 10$ yr for main sequence stars with masses $0.2\le M_{\star}/M_{\odot} \le 5$ and $M_{\bullet} = 10^6M_{\odot}$. The second estimate, which postulates that the star is completely destroyed when tides dominate the maximum stellar self-gravity, predicts that $t_{\rm peak}$ is very weakly dependent on stellar type, {}{with $t_{\rm peak} = \left(23.2\pm4.0\textrm{ days}\right)\left(M_{\bullet}/10^6M_{\odot}\right)^{1/2}$ for $0.2\le M_{\star}/M_{\odot} \le 5$, while $t_{\rm peak} = \left(29.8\pm3.6\textrm{ days}\right)\left(M_{\bullet}/10^6M_{\odot}\right)^{1/2}$ for a Kroupa initial mass function truncated at $1.5 M_{\odot}$}. This second estimate also agrees closely with hydrodynamical simulations, while the frozen-in model is discrepant by orders of magnitude. We conclude that (1) the time to peak luminosity in complete TDEs is almost exclusively determined by SMBH mass, and (2) massive-star TDEs power the largest accretion luminosities. Consequently, (a) decades-long extra-galactic outbursts cannot be powered by complete TDEs, including massive-star disruptions, and (b) the most highly super-Eddington TDEs are powered by the complete disruption of massive stars, which -- if responsible for producing jetted TDEs -- would explain the rarity of jetted TDEs and their preference for young and star-forming host galaxies.
\end{abstract}


\section{Introduction}
\label{sec:intro}
In a tidal disruption event (TDE), a star is either completely or partially ripped apart by the tidal field of a supermassive black hole (SMBH) \citep{hills75,rees88, gezari21}. Approximately half of the stripped stellar debris is bound to the black hole, returns to the point of disruption, and circularizes to form an accretion disc \citep{hayasaki13,Shiokawa15,Bonnerot16,sadowski16,curd19,andalman22}. The process of accretion generates an electromagnetic flare, the peak luminosity of which can be comparable to the Eddington limit of the SMBH \citep{evans89,wu18}, and many of these flares have been observed recently with the advent of survey science (e.g., \citealt{nicholl19, pasham19, wevers19, hinkle21, vanvelzen21, payne21, wevers21, lin22, nicholl22, hammerstein23, pasham23, wevers23, yao23}; see also \citealt{gezari21} and references therein).

The process of TDE disc formation is still not well understood. However, the fallback rate, being the rate at which stellar debris returns to pericenter (denoted by $\dot{M}$), is the main factor that determines the disc properties (at least for times $\lesssim$ few years, after which viscous delays could yield substantial differences between the fallback and accretion rates; \citealt{cannizzo90}). Considerable effort has thus been dedicated to modeling the fallback of debris from a TDE, the earliest analytic model of which, known as the frozen-in approximation \citep{lacy82,bicknell83,stone13}, assumes that the star is completely destroyed by the SMBH and the binding energy of the debris (to the SMBH) is established at the tidal radius, $r_{\rm t}$. The tidal radius is determined by equating the tidal acceleration to the surface gravitational field of the star, and is given by 
\begin{equation}
\label{tidalradius}
    r_{\rm t}=R_{\star} \left( M_{\bullet}/M_{\star} \right) ^{1/3},
\end{equation}
where $M_{\bullet}$ is the mass of the SMBH and $R_{\star}$ and $M_{\star}$ are the stellar radius and mass, respectively. Using the frozen-in model, \cite{lodato09} demonstrated that the peak fallback rate and the time to peak are strongly dependent on stellar structure and, thus, the stellar mass and age.

Numerical simulations of the disruption of a $5/3$-polytropic star have found reasonably good agreement with the frozen-in model (e.g., \citealt{lacy82, evans89, lodato09}), but simulations of the disruption of stars with more realistic stellar structure have uncovered notable differences between the two. For example, \citet{golightly19} found that the disruption of a $1 M_{\odot}$ star at zero-age main sequence (evolved with the stellar evolution code {\sc mesa}; \citealt{paxton11}) yielded a peak fallback time of $\sim 25$ days, as compared to the frozen-in prediction of $\sim 1$ yr (see Figure 2 in \citealt{golightly19}). \cite{coughlin22} suggested that this discrepancy is at least partially related to the definition of the tidal radius, and in particular they noted that a star should be completely destroyed when the tidal field of the black hole exceeds the maximum self-gravitational field within the star, not the surface gravitational field (as the latter is always less than the former). From their model (hereafter the CN22 model), they showed that the time to the peak in the fallback rate could be calculated from the properties of the star, and found good agreement between their predictions and numerical hydrodynamical simulations presented in~\cite{guillochon13,golightly19,nixon21}. The question is then: does the CN22 predict a strong dependence between the time to peak fallback rate and stellar type?

Here we show that the answer to this question is no: in Section~\ref{sec:peakfbrs} we use the CN22 model to calculate the peak fallback time for stars with zero-age main sequence (ZAMS) masses between $0.2 - 5 M_{\odot}$ at various ages, and we show that the time to peak is $t_{\rm peak} = \left(23.2 \pm 4.0\right)\times\left(M_{\bullet}/10^6M_{\odot}\right)^{1/2}$ days, {}{while $t_{\rm peak} = \left(29.8 \pm 3.6\right)\times\left(M_{\bullet}/10^6M_{\odot}\right)^{1/2}$ days if the population is weighted by a Kroupa initial mass function (IMF) truncated at $1.5\, \Msun$} \citep{kroupa01,stone16}. We also show that these predictions are in very good agreement with numerical simulations, and we thus conclude that the time to peak in a complete disruption is almost exclusively determined by SMBH mass. 

We discuss the implications of our findings in Section \ref{sec:conclusion}, including the role of partial disruptions and the production of super-Eddington TDEs. In particular, we show that the most highly super-Eddington (by factors $\gtrsim 1000$ for a $10^6M_{\odot}$ SMBH) TDEs are generated when the destroyed star is of high mass, and hence jetted TDEs -- which may be produced as a byproduct of hypercritical accretion -- could be most easily generated through the disruption of massive (and rare) stars. We also show that the complete disruption of high-mass stars cannot yield long-duration ($\gtrsim 10$ yrs) transients, as has been invoked in some investigations, and instead can only be consistent with observations if the disruption was partial. We summarize our findings in Section \ref{sec:summary}.

\section{Peak Fallback Rates}
\label{sec:peakfbrs}

\begin{figure*}[t]
    \advance\leftskip-0.5cm
    \includegraphics[height=6cm,width=0.53\textwidth]{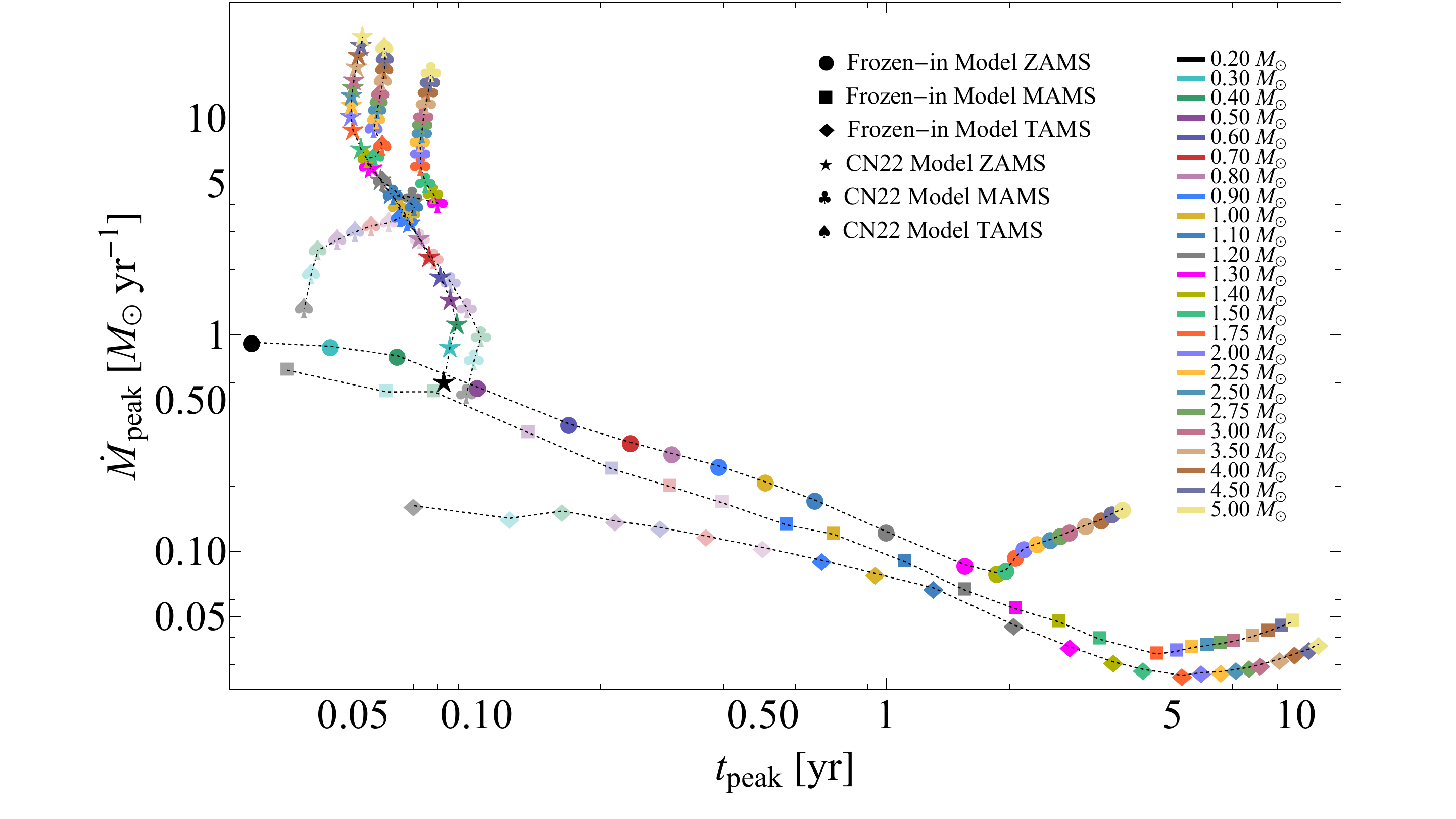}\hfill
    \includegraphics[height=6cm,width=0.53\textwidth]{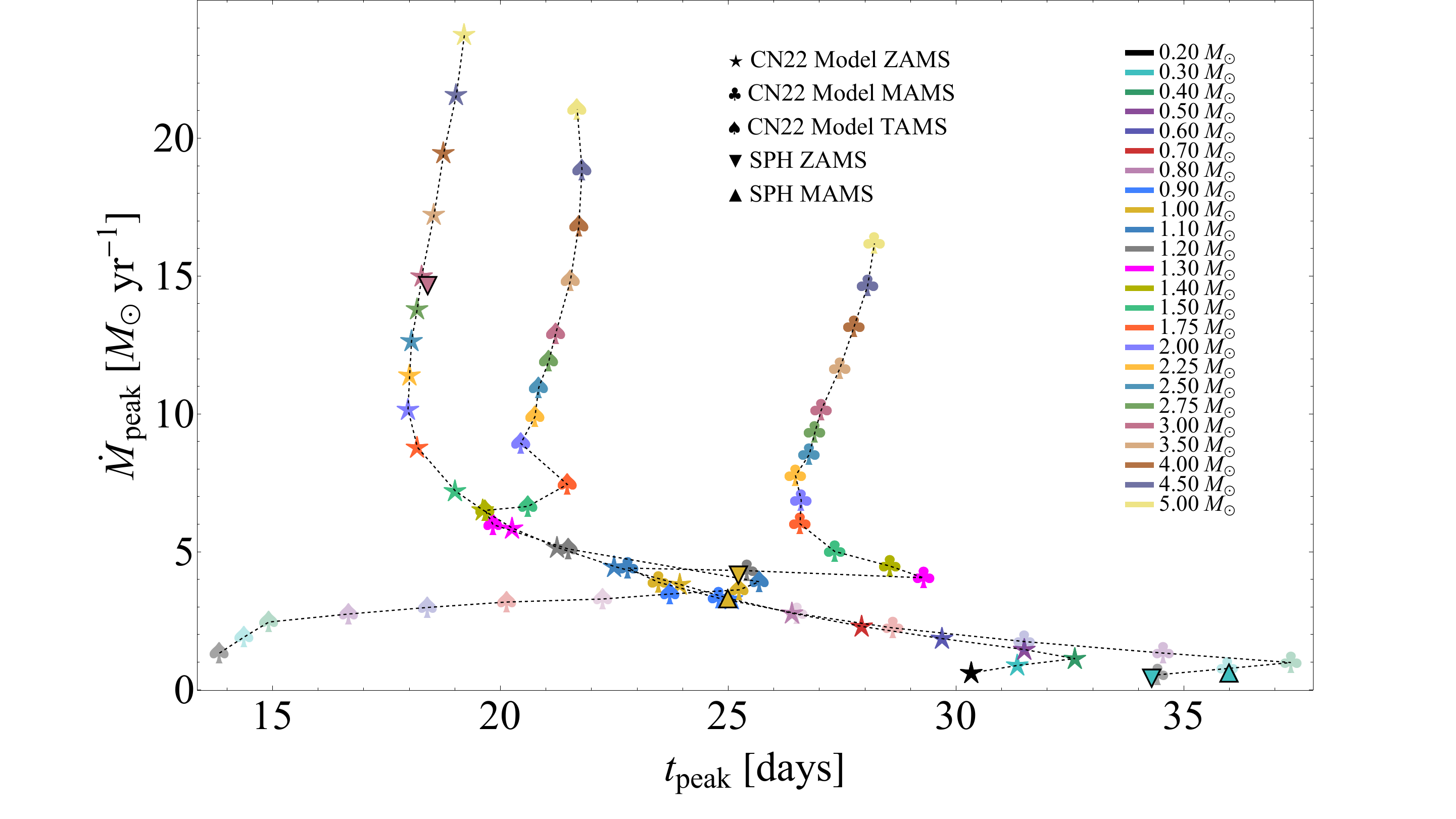}
    \caption{Left: Peak fallback times, $t_{\rm peak}$, and rates, $\dot{M}_{\rm peak}$, for stars disrupted by a $10^6M_{\odot}$ SMBH using the frozen-in and CN22 models; the stellar mass and age are indicated in the legends (the black, dashed lines join stars of a given age). Right: Same as the left panel, but showing only the predictions of the CN22 model (note $t_{\rm peak}$ is in days) and numerical hydrodynamics simulations (shown by the inverted and upright triangles). (Some low-mass stars reach neither MAMS nor TAMS in 14 Gyr; we nonetheless include these data points to illustrate the predictions of the model over many stellar types, but their coloration is semi-transparent).}
    \label{frozenin-cn22-fbr}
\end{figure*}

Using the stellar evolution code {\sc mesa}~\citep{paxton11}, we evolved solar-metallicity (at zero-age main sequence; ZAMS) stars in the mass range $0.2-5\Msun$ from the ZAMS to the terminal-age main sequence (TAMS), when the central hydrogen mass fraction drops below $0.1\%$. We then calculated the peak fallback rates associated with the disruption of these stars by a $10^6 M_{\odot}$ SMBH with two different methods, the first of which uses the frozen-in approximation. The frozen-in model assumes the star retains perfect hydrostatic balance until reaching $r_{\rm t}$, after which point the fluid elements comprising the original star trace out independent, Keplerian orbits in the SMBH's gravitational field. The fallback rate of the bound debris -- accounting for the structure of the star -- is then determined using the formalism described in \cite{lodato09}, wherein vertical ``slices'' of the star return contemporaneously to the SMBH. 

In the second method we adopted the CN22 model, which postulates that the star is completely destroyed if the tidal field of the SMBH exceeds the maximum self-gravitational field within the star. The radius within the star at which the self-gravitational field is maximized is defined as the core radius $R_{\rm c}$, and equating the self-gravitational field at $R_{\rm c}$ to the tidal field (at $R_{\rm c}$) then yields the core disruption radius $r_{\rm t,c}$. Assuming that the core (i.e., every fluid element at radii $\le R_{\rm c}$ within the star) retains approximate hydrostatic balance until reaching $r_{\rm t, c}$, the time to peak and the peak value of the fallback rate are (see Equations 11 and 12 in \citealt{coughlin22} and their discussion)
\begin{equation}
    \label{tpeak}
    t_{\rm peak} = \left( \frac{r_{\rm t,c}^2}{2 R_{\rm c}}\right)^{3/2} \frac{\pi}{\sqrt{G M_{\bullet}}}, \quad \dot{M}_{\rm peak} = \frac{M_{\star}}{4 t_{\rm peak}}.
\end{equation}

\begin{figure}[t]
\advance\leftskip-0.5cm
\includegraphics[width=0.52\textwidth]{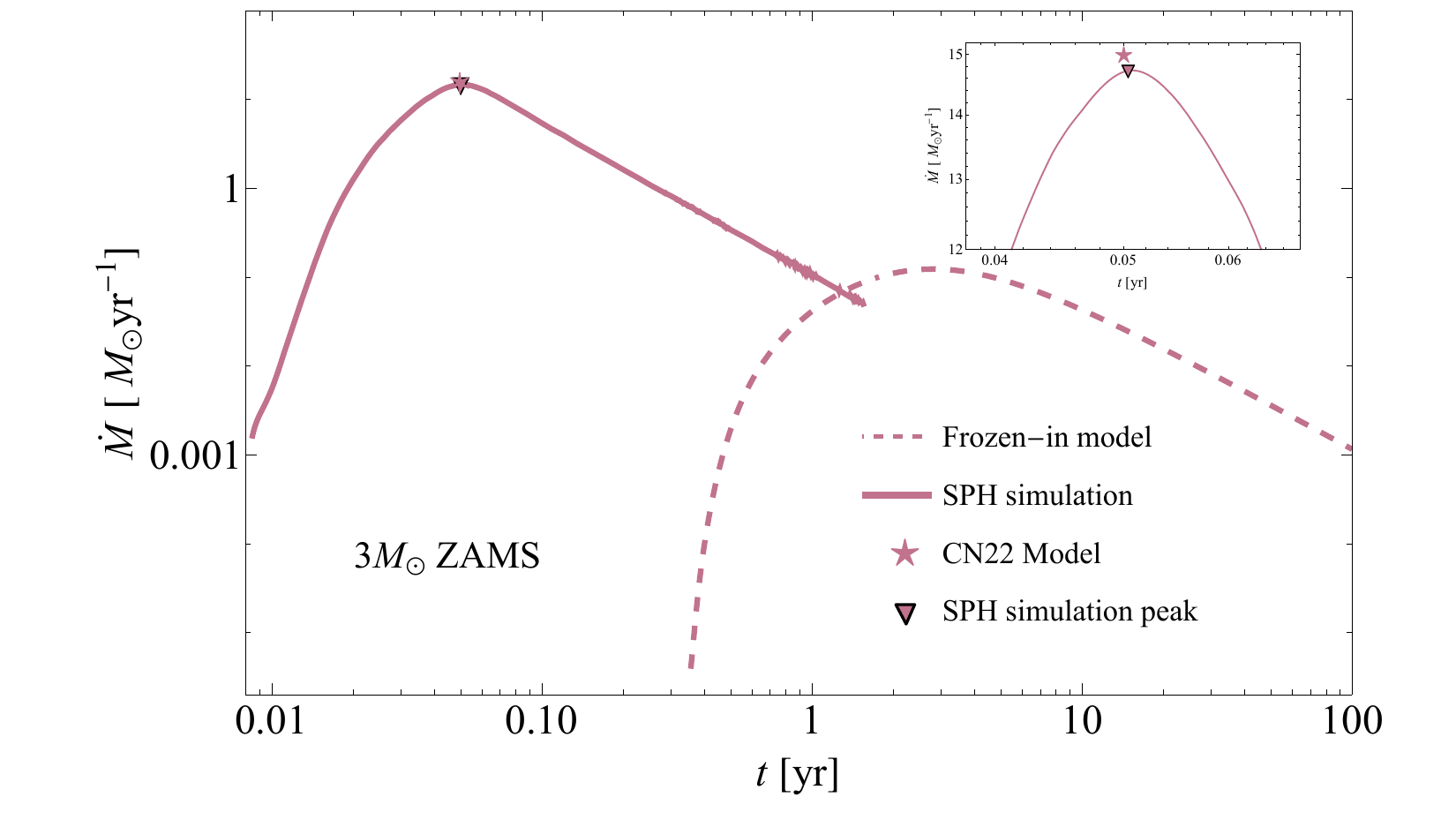}
\caption{The fallback rate for the complete disruption of a $3\Msun$ ZAMS star by a $10^6 \Msun$ SMBH. The dashed curve shows the frozen-in approximation following the prescription in \cite{lodato09}, while the solid curve is from a hydrodynamical simulation. The inset shows a zoomed-in view near the peak.} 
\label{3msun-zams-fbr}
\end{figure}

\begin{figure*}[t]
    \includegraphics[height=6cm,width=0.495\linewidth]{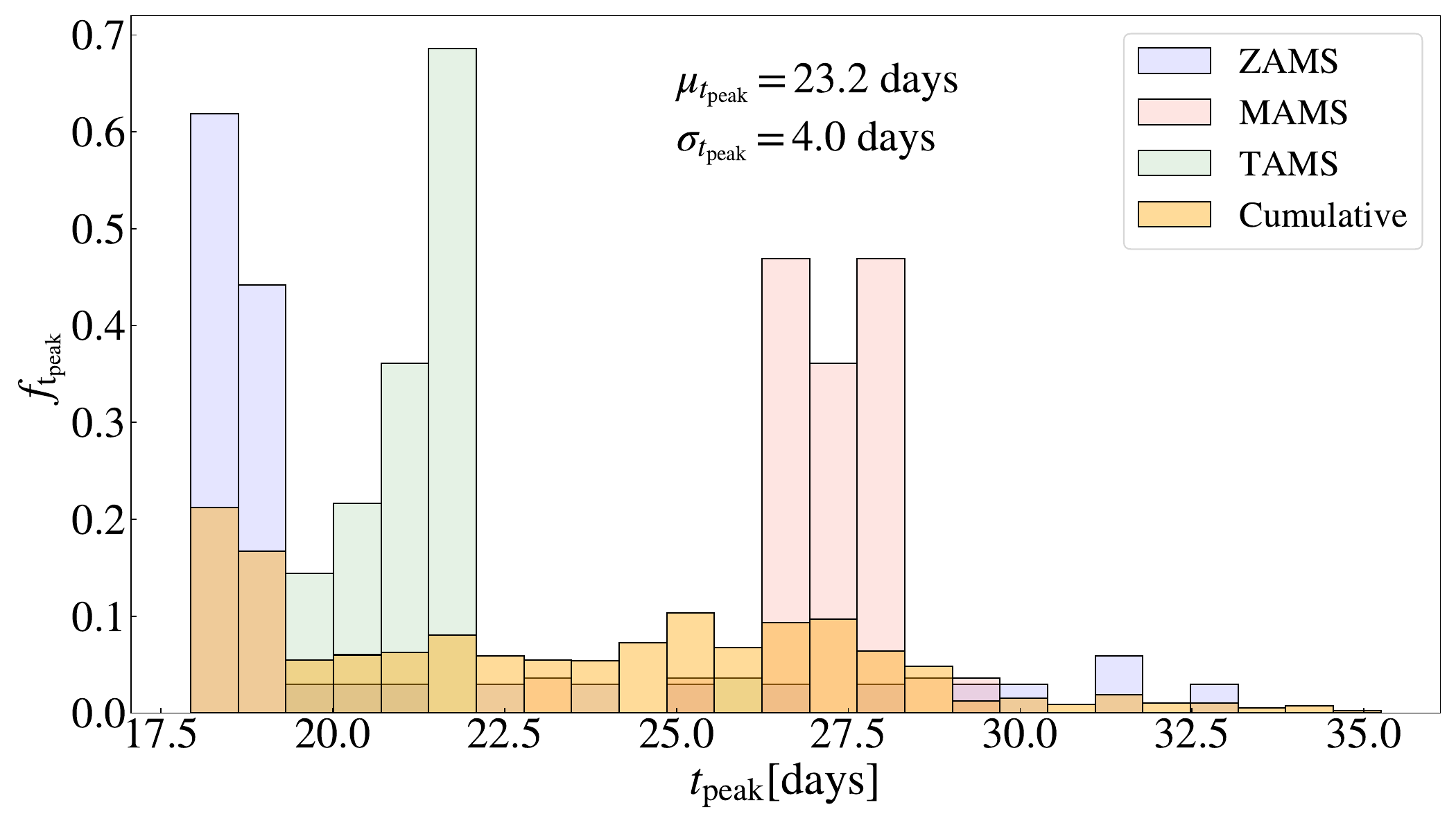}\hfill
    \includegraphics[height=6cm,width=0.495\linewidth]{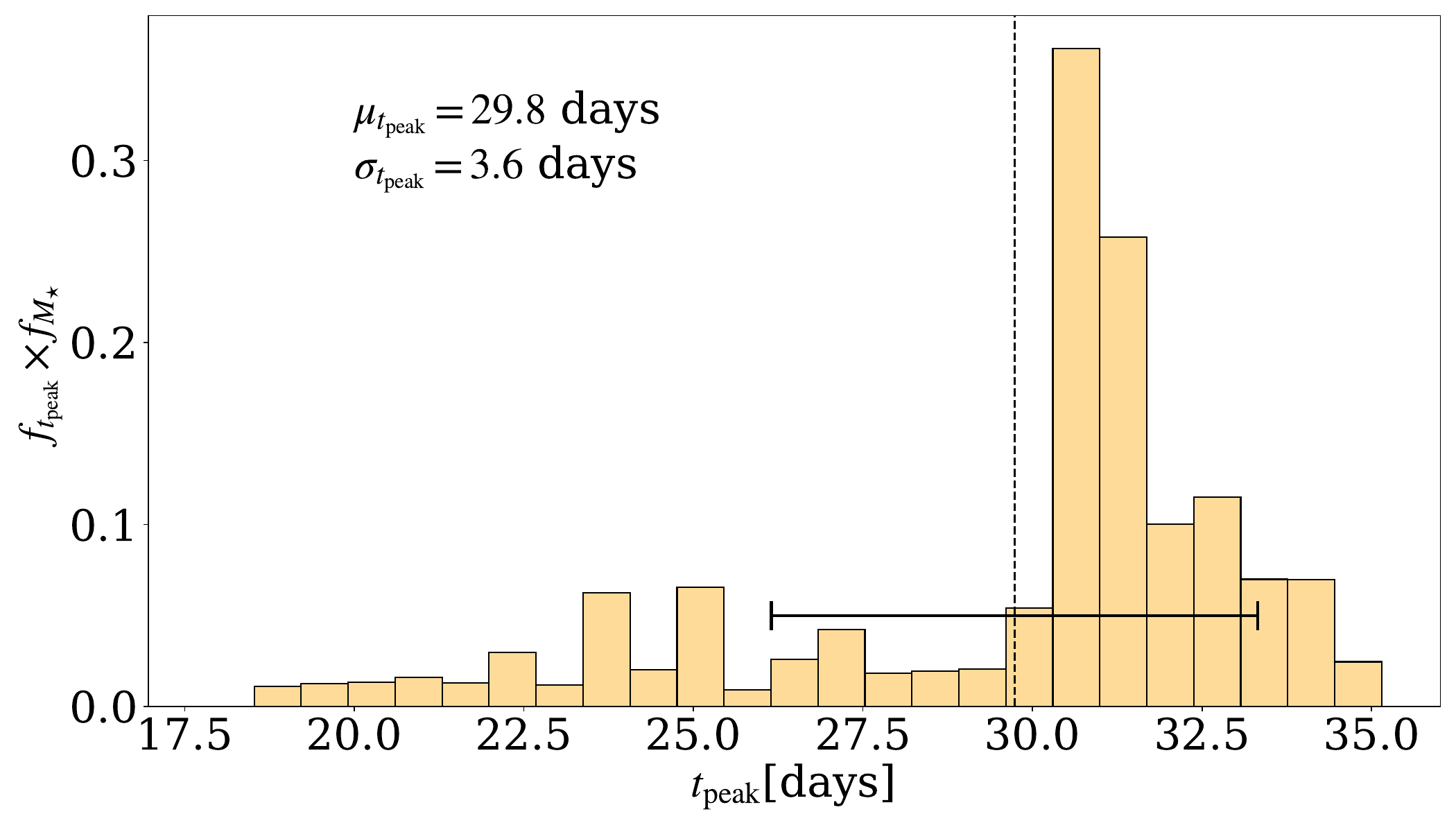}
    \caption{{Left: The probability distribution function of peak times, $f_{t_{\rm peak}}$, for the tidal disruption of $0.2-5\Msun$ stars (at different ages in the main sequence) by a $10^6\Msun$ SMBH. The mean and standard deviation of the distribution for all stars is given by $t_{\rm peak}=23.2\pm4.0$ days. Right: $f_{t_{\rm peak}}$ weighted by a {}{Kroupa initial mass function truncated at $1.5 \, \Msun$}, to account for the observed over-abundance of low mass stars in the universe, giving $t_{\rm peak}=29.8\pm3.6$ days. The vertical dashed line indicates the mean of the probability distribution, $\mu_{\rm t_{\rm peak}}$, and the $1\sigma$ confidence interval, $\sigma_{\rm t_{\rm peak}}$, is indicated by the horizontal errorbar.}}
    \label{tpeak_histogram}
\end{figure*}

The left panel of Figure~\ref{frozenin-cn22-fbr} shows $\dot{M}_{\rm peak}$ and $t_{\rm peak}$ for the ZAMS masses given in the legend (and a $10^6M_{\odot}$ SMBH) and at three ages -- ZAMS, TAMS, and when the central hydrogen mass fraction falls below 0.2, the ``middle-age main sequence'' (MAMS) -- for both the frozen-in method and the CN22 model. The core disruption model of CN22 predicts that $t_{\rm peak}$ varies between $\sim 18-33$ days for all stars that reach their given age within a Hubble time\footnote{Stars below $\sim 0.9 \Msun$ do not reach MAMS or TAMS within $14$ Gyr and have thus been colored semi-transparently.}. On the other hand, the frozen-in approximation predicts a peak fallback time that depends strongly on the type of star, and spans $\sim 2$ orders of magnitude (from $\sim 10$ days to $\sim few-10$ years) as the stellar mass varies from $0.2$ to $5M_{\odot}$. $\dot{M}_{\rm peak}$ also differs greatly between the two models: because $t_{\rm peak}$ is nearly independent of stellar type for the CN22 model, $\dot{M}_{\rm peak}$ scales roughly in proportion with the stellar mass, while the frozen-in $\dot{M}_{\rm peak}$ declines with mass until $M_{\star} \simeq 2 M_{\odot}$. Therefore, while the CN22 model predicts a peak fallback rate of $\sim20 M_{\odot}$ yr$^{-1}$ for a $5 M_{\odot}$ TAMS star, the frozen-in model predicts $\dot{M}_{\rm peak} \sim 0.04 M_{\odot}$ yr$^{-1}$, i.e., smaller by nearly three orders of magnitude. 

To investigate the validity of the CN22 model, we compared its predictions to hydrodynamical simulations, specifically those described in \citet{golightly19,nixon21}. The right panel of Figure~\ref{frozenin-cn22-fbr} shows the peak fallback rates and peak times (now in days) obtained using the CN22 core disruption model on a linear-linear scale. For the $3M_{\odot}$ ZAMS star the SPH data point is $(18.4\,\,\mathrm{days},14.7\Msun \rm yr^{-1})$, the $1M_{\odot}$ ZAMS and MAMS data points are $(25.2\,\,\mathrm{days},4.2\Msun \rm yr^{-1})$ and $(25.0\,\,\mathrm{days},3.4\Msun \rm yr^{-1})$, and the $0.3 M_{\odot}$ ZAMS and MAMS data points are $(34.3\,\,\mathrm{days},0.5\Msun \rm yr^{-1})$ and $(36.0\,\,\mathrm{days},0.7\Msun \rm yr^{-1})$ respectively. This demonstrates excellent agreement between the model and the numerical results\footnote{Note that the simulations in \citet{golightly19}, from which these data points were taken, were all performed with $\beta = 3$. As $\beta_{\rm c}$ is a function of stellar mass (see Figure \ref{betac}), the ratio of $\beta/\beta_{\rm c}$ is largest for $0.3M_{\odot}$ ZAMS, and this data point shows the largest discrepancy with the CN22 model (perhaps indicating that there is a weak dependence of the peak fallback rate on $\beta > \beta_{\rm c}$, consistent with \citealt{guillochon13}). Note also that the $3M_{\odot}$ MAMS disruption and all of the TAMS disruptions performed in \citet{golightly19} were only partial -- which is also consistent with Figure \ref{betac} -- and hence there are no numerical data points of complete disruptions for these stars; $1M_{\odot}$ MAMS was also a partial disruption, but only a relatively small core survived and the peak rate and time is insensitive to $\beta$ at $\beta \gtrsim 2$ for this star (see Figure 4 of \citealt{nixon21}).}. To further highlight the differences among these models, Figure~\ref{3msun-zams-fbr} compares the fallback rate from the complete disruption of a $3\Msun$ star at ZAMS, computed using the frozen-in approximation, to that obtained from the SPH simulations described in \cite{golightly19} for the same {\sc mesa} star. 

The narrow range of $t_{\rm peak}$ predicted by the CN22 model is further illustrated in Figure~\ref{tpeak_histogram}, the left panel of which shows the probability distribution function of $t_{\rm peak}$ for a population of main-sequence stars in the mass range $0.2-5\Msun$ (i.e., this is a histogram of $t_{\rm peak}$ over this range of stars, with a mass bin size of $0.1M_{\odot}$, normalized over the range of $t_{\rm peak}$). To approximately account for the observational bias that is introduced by the stellar mass function (see Section \ref{sec:tpeak} for additional discussion and caveats related to the stellar mass function), which is heavily weighted toward low-mass stars, we plot the $t_{\rm peak}$ distribution weighted by {}{a present day mass function (PDMF), which is crudely approximated by a Kroupa IMF \citep{kroupa01} truncated at $1.5 \, \Msun$}, in the right panel of Figure~\ref{tpeak_histogram}. The normalized form of the PDMF is given by

\[ f_{M_{\star}}= \left\{
\begin{array}{ll}
      0.50\times \left(M_{\star}/M_{\odot}\right)^{-1.3}, & \,\,\,\, 0.2\leqslant M_{\star}/M_{\odot} < 0.5 \\
      0.25 \times \left(M_{\star}/M_{\odot}\right)^{-2.3}, & \,\,\,\, 0.5\leqslant M_{\star}/M_{\odot}\leqslant 1.5
\end{array} 
\right. \]
We see that the expected range of variation of $t_{\rm peak}$ for the tidal disruption of $0.2-5\Msun$ main sequence stars by a $10^6\Msun$ SMBH is $29.8 \pm 3.6$ days. 

\section{Conclusions and Implications}
\label{sec:conclusion}
\subsection{The $t_{\rm peak}$ distribution and relation to the stellar mass function}
\label{sec:tpeak}
Figure \ref{frozenin-cn22-fbr} demonstrates that, according to the CN22 model and hydrodynamical simulations (see also Figure \ref{3msun-zams-fbr}), the time to the peak in a TDE fallback rate (if the star is completely destroyed) is very insensitive to both the mass and age of the star. {}{Figure \ref{tpeak_histogram} shows that the resulting mean, $\mu_{\rm t_{\rm peak}}$, and standard deviation, $\sigma_{\rm t_{\rm peak}}$, are $\mu_{\rm t_{\rm peak}} = 29.8$ days and $\sigma_{\rm t_{\rm peak}} = 3.6$ days for a $10^{6}M_{\odot}$ SMBH if the distribution is weighted by the PDMF, derived by truncating the Kroupa IMF at $1.5\, \Msun$}. {}{Weighting by the PDMF is a simplified representation of the effect of the stellar mass function on TDE rates, and very crudely approximately accounts for stellar evolution and ongoing star formation for a given galaxy~\citep{Chabrier03}.} Existing studies of TDE rates have often used the Kroupa IMF truncated at a high mass to estimate the PDMF~\citep{mageshwaran2015,stone16}. For example, \cite{stone16} use the Kroupa IMF, truncated at $1\, \Msun$, to approximate the PDMF for an old stellar population. This is likely to be a reasonable approximation for an early-type galaxy, as in this case a large fraction of the high-mass stars -- which have considerably shorter lifetimes -- will have ended their lives as compact objects, thus causing the stellar mass function to be more heavily dominated by low-mass stars. Consequently, for early-type galaxies the $t_{\rm peak}$ distribution would be extremely tightly peaked around 30 days (see the right panel of Figure \ref{frozenin-cn22-fbr}). {}{For comparison, weighting the $t_{\rm peak}$ distribution by the Kroupa IMF (which would have a relatively higher representation of high mass stars) yields a distribution of $t_{\rm peak}$ values with mean and standard deviation given by $t_{\rm peak}=28.9 \pm 4.3$ days, while a PDMF obtained by truncating the Kroupa IMF at $1 \, \Msun$ gives a distribution that is effectively identical to the distribution obtained by truncating the IMF at $1.5 \, \Msun$.}

However, observational evidence suggests that TDEs tend to occur preferentially in E+A galaxies (or E+A-like; \citealt{arcavi14,french16,french20}), which have recently undergone an epoch of star formation that could be related to a merger (e.g., \citealt{dressler83, zabludoff96}). For such galaxies, there is still a substantial population of massive stars, and the PDMF may correspondingly be closer to a Kroupa (or other) IMF (as pointed out in, e.g., \citealt{zhang2018,toyouchi2022}). In such scenarios, the disruption of stars more massive than $\sim 1.5 \, \Msun$ could be more frequent, causing a somewhat larger spread in the distribution of $t_{\rm peak}$ values (i.e., the peak fallback time of massive stars is systematically shorter than $\sim 30$ days, as shown in the right panel of Figure \ref{frozenin-cn22-fbr}, which will broaden the distribution of peak fallback times to include slightly smaller values).

An accurate calculation of the rate of observed TDEs that incorporates the distribution of stellar masses would necessarily account for other dynamical factors that govern the rate of stellar diffusion into the loss cone, the most popularly agreed upon mechanism for which is two-body relaxation (e.g., \citealt{peebles72, bahcall76, frank76, lightman77, cohn78, magorrian99, merritt13, kochanek16, stone16, stone2020}). For realistic stellar populations, typically, the heaviest surviving species dominates the relaxation rate, making mass segregation another important factor in the determination of TDE rates in galactic nuclei. Mass segregation occurs over a fraction of the two-body relaxation time, and causes more massive stellar objects to aggregate near the galactic nucleus while driving lighter objects to larger distances (e.g., \citealt{merritt13,generozov22}). For stellar distributions that include massive stars, e.g., the recent starbursts in E+A galaxies, mass segregation could be an additional factor that results in an over-representation of high mass stars in the rates of observed TDEs.

\subsection{The $t_{\rm peak}$ distribution and dependence on black hole mass}
\label{sec:blackhole}
The CN22 model predicts (as does the frozen-in approximation) that the fallback time varies with SMBH mass as $M_{\bullet}^{1/2}$, which has been verified to be highly accurate with hydrodynamical simulations \citep{wu18}. Therefore, the time to peak for complete disruptions for a given star and black hole mass is
\begin{equation}
    t_{\rm peak} = \left(29.8\pm 3.6\textrm{ days}\right)\times\left(\frac{M_{\bullet}}{10^6M_{\odot}}\right)^{1/2}, \label{tpeakofM}
\end{equation}
suggesting that the time to the peak fallback rate is determined almost exclusively by the SMBH mass for complete disruptions. 

If the TDE luminosity tracks the rate of return of the debris to the SMBH, which comparisons between observations and models suggests is the case \citep{mockler19, nicholl22}, the time to reach the peak luminosity in a TDE -- barring partial disruptions -- should be determined largely by the black hole mass (see also \citealt{guillochon13,mockler19}). Furthermore, because the scaling of $t_{\rm peak}$ with SMBH mass is not strong and because low-mass stars cannot be completely destroyed by non-spinning SMBHs with mass $\gtrsim few\times 10^{7} M_{\odot}$ (while rapid black hole spin increases this upper limit, it is still very unlikely for the star to be destroyed and not directly captured for SMBHs with mass above $\sim few\times 10^{7}M_{\odot}$, independent of the black hole spin; see Figure 6 in \citealt{coughlin22b}), we would expect the majority of TDEs to have times to peak that satisfy $t_{\rm peak} \lesssim 100$ days. There is some ambiguity in making direct comparisons between these predictions and observations, as our $t_{\rm peak}$ is measured from the time to pericenter, and the time of disruption is not directly measurable for TDEs\footnote{There are two potential exceptions to this: 1) The ``recombination transient,'' as described in \citet{kasen10}, which occurs as ionized hydrogen recombines, could illuminate the otherwise-dark period between the initial disruption and the return of the most-bound debris, but this was recently shown by \citet{coughlin23} to be much dimmer than originally predicted; and 2) In a repeating partial TDE, where the star is bound to the SMBH and stripped of mass at each pericenter passage \citep{payne21}, the return of the star to pericenter results in a sharp decline in the luminosity if the fallback and accretion rates are tightly coupled \citep{wevers23}. The time between the cutoff in the emission and the rebrightening is then a direct measure of the fallback time, as described in \citet{wevers23} for the specific case of AT2018fyk \citep{wevers19}.}. However, the 30 optically bright TDEs detected by the Zwicky Transient Facility \citep{bellm19}, as described in \citet{hammerstein23}, all exhibit times to peak (from first detection) that are clustered around $\sim 50$ days (see Figures 17 and 18 of \citealt{hammerstein23}). 

\subsection{The impact of partial disruptions}
\label{sec:partial}
In addition to the SMBH mass, the time to peak in the fallback rate is also modified if the disruption is only partial \citep{guillochon13}. In particular, as the pericenter of the star increases beyond the distance necessary for complete disruption, the peak fallback rate declines and the time to peak extends to later times (e.g., Figure 5 of \citealt{guillochon13}, Figure 8 of \citealt{gafton19}, Figure 7 of \citealt{lawsmith20}, and Figures 1 and 2 of \citealt{nixon21}). This shift in the peak time therefore results in a degeneracy between the black hole mass and the pericenter distance of the star. However, the steeper decline of the fallback rate for a partial TDE \citep{guillochon13}, which scales as $\propto t^{-9/4}$ in particular \citep{coughlin19, golightly19, miles20, nixon21, wang21, kremer23}, should, in principle, offer a means to disambiguate these two effects.

\begin{figure*}
\includegraphics[height=5.2cm,width=0.495\textwidth]{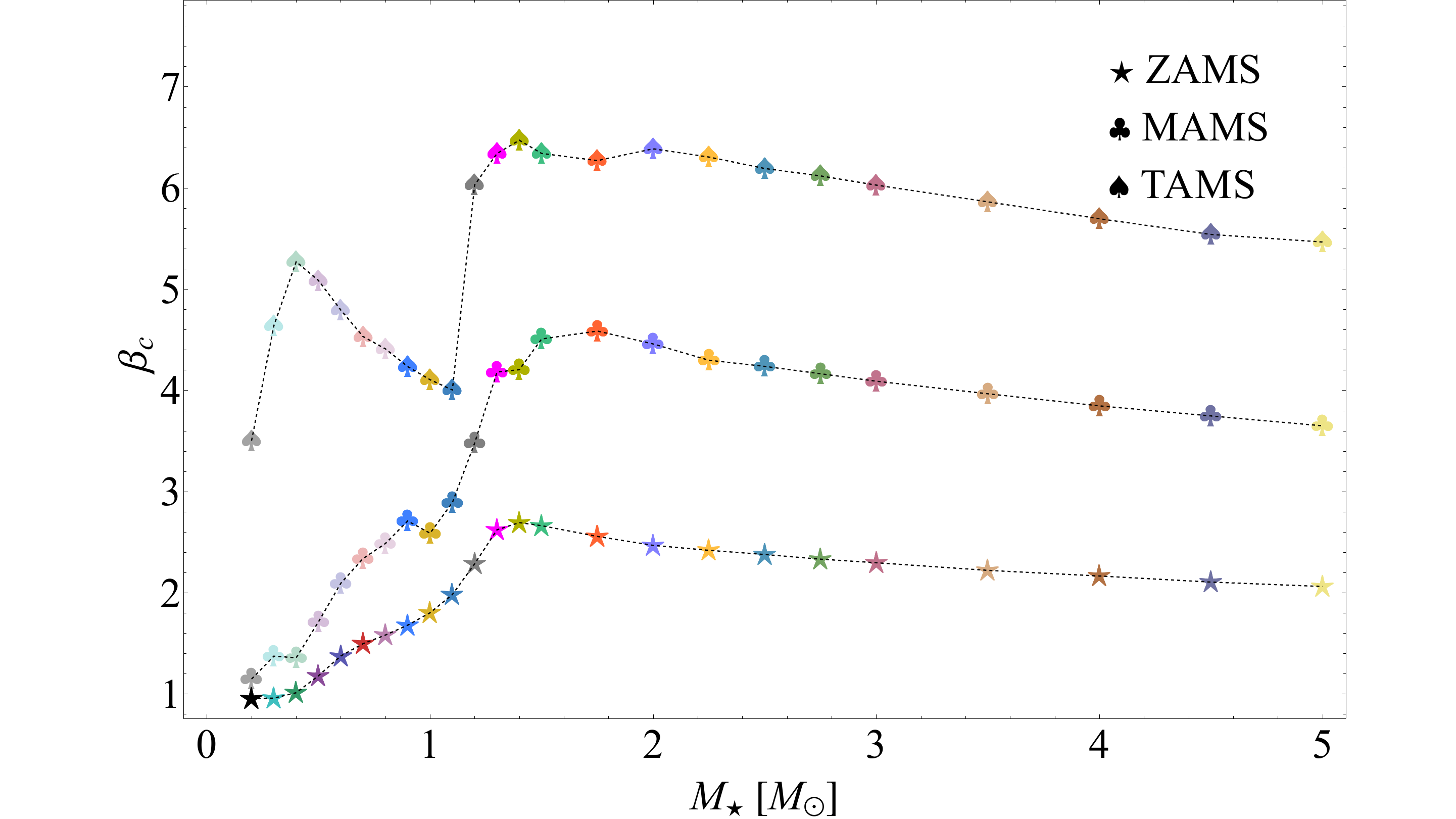}
\includegraphics[height=5.2cm,width=0.495\textwidth]{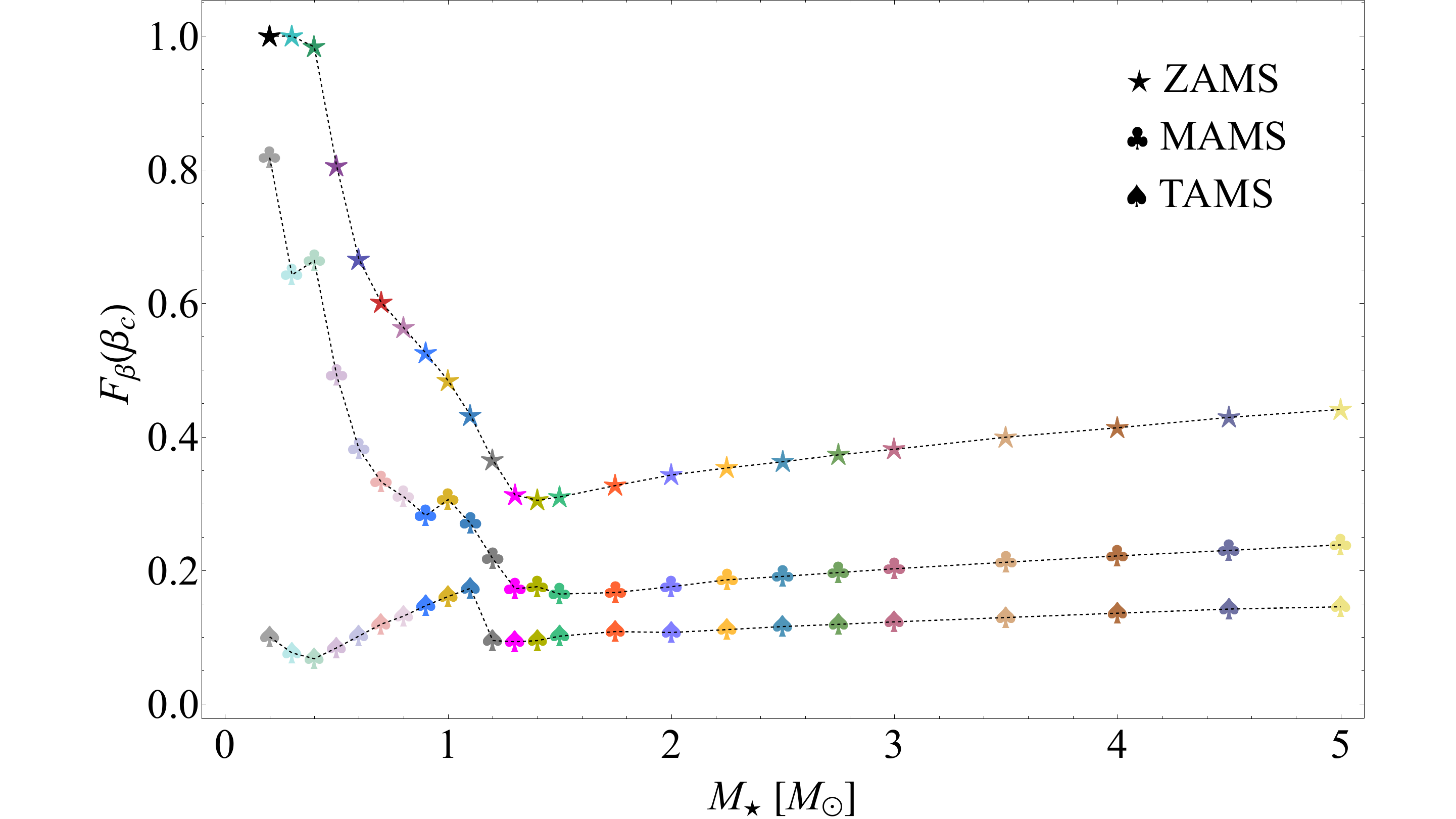}
    \caption{Left: The impact parameter $\beta_{\rm c} = r_{\rm t}/r_{\rm t, c}$ required for complete disruption. As stellar age increases the star becomes more centrally concentrated, resulting in larger $\beta_{\rm c}$. Right: The fraction of stars that enter within the fiducial tidal radius $r_{\rm t}$, have impact parameters $\beta > \beta_{\rm c}$ (i.e., are completely destroyed), and are not directly captured by a Schwarzschild SMBH with $M_{\bullet} = 10^6M_{\odot}$.  }
    \label{betac}
\end{figure*}

The possibility of partial disruption, and the likelihood of the increased peak time as a consequence of this effect, is particularly relevant for evolved stars, as the increasing central stellar density as the star ages results in a reduction in the core tidal disruption radius and makes complete disruptions less likely. This feature is shown in Figure \ref{betac}, the left panel of which gives the impact parameter $\beta_{\rm c} \equiv r_{\rm t}/r_{\rm t, c}$, i.e., the inverse of the ratio of the core tidal disruption radius to the fiducial tidal radius, as determined from the CN22 model for the same population of stars in Figure \ref{frozenin-cn22-fbr}. Larger values imply that the star must penetrate to smaller radii to be completely disrupted, which is statistically less likely or impossible without being directly captured. For example, in the case of a Schwarzschild SMBH, the cumulative distribution function of $\beta$ (i.e., the number of TDEs, which are stars that come within $r_{\rm t}$ but outside the direct capture radius, with impact parameter greater than $\beta$) in the pinhole regime is \citep{coughlin22b}
\begin{equation}
    F_{\beta}(\beta) = \frac{\left(r_{\rm t}-2R_{\rm g}\right)\left(r_{\rm t}-4\beta R_{\rm g}\right)^2}{\beta\left(r_{\rm t}-2\beta R_{\rm g}\right)\left(r_{\rm t}-4R_{\rm g}\right)^2},
\end{equation}
where $R_{\rm g} = GM_{\bullet}/c^2$. This expression assumes that the distribution function of stars in angular momentum space has reached a steady-state and that the loss cone is full, which necessitates that stars re-populate the loss cone at a rate that is sufficient to maintain a constant rate of consumption by the SMBH (see the references listed in the last paragraph of Section \ref{sec:tpeak}). It may be the case that, especially for galaxies in which the TDE rate is high (e.g., the E+A population referenced in Section \ref{sec:tpeak}), this assumption is not valid, requiring a more detailed analysis of the time dependence of the distribution function. This analysis is beyond the scope of the present work.

The right panel of Figure \ref{betac} shows $F_{\beta}$ for a $10^6 M_{\odot}$ SMBH as a function of stellar mass for the three different stellar ages, and highlights the fact that the fraction of complete disruptions\footnote{Note that the largest value of $\beta_{\rm c}$ is $\beta_{\rm c} \simeq 6.47$ for a $1.4M_{\odot}$ TAMS star, which is not too close to the direct capture value for a Schwarzschild black hole, being $\beta_{\rm dc} \gtrsim 10$ for the stars considered here, and hence the black hole spin does not significantly modify these results. For example, using the formalism given in \citet{coughlin22b}, $F_{\beta}(6.47) \simeq 0.0937$ for black hole spin $a = 0$, while $F_{\beta}(6.47) \simeq 0.104$ for $a = 0.999$.} can be $\lesssim 10\%$ for stars that have $\beta_{\rm c} \gtrsim 6$. This fraction declines substantially as the SMBH mass increases because of the fact that the gravitational radius grows as $\propto M_{\bullet}$ while the tidal radius scales as $\propto M_{\bullet}^{1/3}$.

\subsection{Long-duration TDEs from the disruption of massive stars}
\label{sec:massive}
The standard scaling of fallback rate predicted by the frozen-in approximation includes a strong dependence on stellar type, and implies that massive stars produce peak fallback timescales that are on the order of years, as is shown in Figure~\ref{frozenin-cn22-fbr}. The corollary of this increased fallback time is that the peak fallback rate is also small, despite the increase in the mass of the disrupted star, which is also apparent from Figure \ref{frozenin-cn22-fbr}. This result -- that the time to peak in the fallback rate is an increasing function of stellar mass -- has been invoked to explain long-duration flares of some TDE candidates. For example, the $\sim$ decades-long X-ray outburst from the galaxy GSN 069, which was recently found to exhibit quasi-periodic eruptions \citep{miniutti19, miniutti23b, miniutti23}, was argued to be due to the disruption of a massive star~\citep{sheng21}. Similar claims in the context of other putative TDEs have been made by \cite{lin17,yan18,subrayan23, cao23}. In \cite{lin17} in particular, the authors note that the disruption of a $10 \, \Msun$ star can explain a long-duration X-ray flare lasting over $\sim 11 $ years. However, this conclusion is only reached by extrapolating the classic fallback rate derived from the frozen-in approximation to massive stars, which Figure \ref{frozenin-cn22-fbr} shows is incorrect for complete disruptions.

Our results demonstrate that these interpretations can only be valid, i.e., that the observed X-ray flares were powered by the disruption of a massive star (and not that the black hole was of a much larger mass, which Equation \ref{tpeakofM} shows could also extend the fallback time), if the disruption was partial. Specifically, since the partial tidal disruption radius -- which is where any mass is successfully stripped from the star -- should coincide with roughly the canonical tidal disruption radius (see the discussion in \citealt{coughlin22}), we would expect massive stars to produce a large range of peak fallback times in going from very little mass lost to complete disruption, and hence relatively long-duration flares could be produced by the partial disruption of massive stars (see \citealt{macleod13}). The TDE powering GSN 069 was recently suggested by \cite{miniutti23} to be a partial, who argued specifically that the decline of the X-ray lightcurve is consistent with a $\propto t^{-9/4}$ decay, and our results provide evidence to support this claim.

\subsection{Super-Eddington accretion and jetted TDEs}
\label{sec:jetted}
Finally, these results have important implications for the number and progenitors of jetted TDEs, which have been argued to arise from super-Eddington accretion onto the black hole. With a radiative efficiency $\epsilon=0.1$ and an (electron scattering) opacity of $\kappa = 0.34$ cm$^{2}$ g$^{-1}$, the fallback rate required for super-Eddington accretion onto a $10^6M_{\odot}$ SMBH is $\dot{M} \approx 0.02 \, \rm \Msun yr^{-1}$. From Figure~\ref{frozenin-cn22-fbr}, we see that the CN22 model predicts peak luminosities that are super-Eddington by at least $\sim 1.5$ orders of magnitude for every star, irrespective of mass or age, and nearly three orders of magnitude for the most massive stars. The frozen-in approximation, on the other hand, predicts lower peak accretion rates as the stellar mass increases (although the peak accretion rate increases slightly as the stellar mass exceeds $\sim 2 M_{\odot}$), with near-Eddington accretion rates for stellar masses $M_{\star} \gtrsim 1.5 M_{\odot}$ at MAMS and TAMS. Thus, using the frozen-in approximation, the only way to generate a highly super-Eddington accretion is to reduce the mass of the black hole. 

It may be that the launching of a relativistic outflow from a TDE requires highly super-Eddington fallback rates, e.g., thousands of times the Eddington limit (as would also be generated under the collapsar paradigm of long gamma-ray bursts; \citealt{woosley93, macfadyen99}). This is consistent with the relativistic and structured outflow solutions of \citet{coughlin20}, which represent jets that are driven and structured by radiation and are matter-dominated (in contrast to Poynting-dominated; such a matter-dominated jet is also consistent with observations of AT2022cmc; \citealt{pasham23}). Specifically, for these solutions the Lorentz factor of the outflow is related to the Eddington factor (i.e., the ratio of the accretion rate to the Eddington luminosity) $\ell$ as $\Gamma \simeq \ell/10$ if the comoving radiation energy density is in rough equipartition with the rest-mass energy density (see Equations 36 -- 38 in \citealt{coughlin20}), and hence to achieve Lorentz factors $\gtrsim 100$ would necessitate Eddington ratios $\gtrsim 1000$.

Our results then show that the relativistic TDEs \emph{Swift} J1644 \citep{bloom11, levan11, zauderer11}, \emph{Swift} J2058 \citep{cenko12}, \emph{Swift} J1112 \citep{brown15}, and AT2022cmc \citep{andreoni22, pasham23} could have been most readily powered by the disruption of a massive star, and the rarity of jetted TDEs could be construed as a consequence of the rarity of massive stars. Additionally, if the disruption of a massive star is necessary to power relativistic TDEs, then one would expect the host galaxies of jetted TDEs to show evidence of recent star formation (analogous arguments, and particularly the dearth of long gamma-ray bursts with early-type galaxy hosts, serve to solidify the supernova gamma-ray burst connection; \citealt{woosley06} and references therein). The host galaxy of the recently detected, jetted TDE AT2022cmc has been inferred to be similar to those from which long gamma-ray bursts usually arise, i.e., the host of AT2022cmc is likely young and star-forming \citep{andreoni22}, as was the host galaxy for Swift J1644+57 \citep{bloom11}.

\section{Summary}
\label{sec:summary}
Using a new model for the peak fallback rate and the time to the peak in a TDE, we have shown that the time to peak is almost completely independent of stellar type (mass or age), which agrees with numerical hydrodynamical simulations and disagrees with the standard, frozen-in model. The main implications of our results are the following:

\begin{enumerate}
    \item{The time to the peak in the fallback rate from a TDE in which the star is completely destroyed is insensitive to stellar type, and -- for stellar mass functions that are dominated by low-mass stars -- is peaked around $30\times\left(M_{\bullet}/(10^6M_{\odot})\right)^{1/2}$ days.} 
    \item{Only increasing the black hole mass or partially disrupting the star can lead to substantially longer fallback times; the complete disruption of a massive star cannot explain the decades-long outbursts that have been observed from some galactic nuclei.}
    \item{The peak accretion rate scales with the mass of the star, implying that the highest-mass stars generate the most highly super-Eddington fallback rates (which contradicts the prediction that would arise from the standard frozen-in approximation). It may therefore be that massive-star disruptions are required to power jetted TDEs, and jetted TDEs are therefore rare because of the rarity of massive stars.}
\end{enumerate}

\begin{acknowledgements}
\section*{}
We thank the referee for comments and suggestions that added to the quality and breadth of our work, and E.R.C.~thanks Thomas Wevers and D.J.~Pasham for useful discussions. E.R.C.~acknowledges support from the National Science Foundation through grant AST-2006684, and from the Oakridge Associated Universities through a Ralph E.~Powe Junior Faculty Enhancement award. This research was supported in part by two MPS-High supplements from the National Science Foundation, which provided stipendiary resources for the Syracuse University Research in Physics (SURPh) program during the summers of 2022 and 2023. CJN acknowledges support from the Science and Technology Facilities Council (grant number ST/Y000544/1) and from the Leverhulme Trust (grant number RPG-2021-380). Part of this research made use of the DiRAC Data Intensive service at Leicester, operated by the University of Leicester IT Services, which forms part of the STFC DiRAC HPC Facility (www.dirac.ac.uk). The equipment was funded by BEIS capital funding via STFC capital grants ST/K000373/1 and ST/R002363/1 and STFC DiRAC Operations grant ST/R001014/1. DiRAC is part of the National e-Infrastructure.
\end{acknowledgements}

\end{document}